# Rheological Behavior of Aqueous Suspensions of Laponite: New Insights into the Ageing Phenomena[§]

Yogesh M Joshi[*], G. Ranjith K. Reddy, Ajit L. Kulkarni, Nishant Kumar,
Raj P. Chhabra

Department of Chemical Engineering, Indian Institute of Technology Kanpur,
Kanpur 208016, INDIA.

[*] Corresponding Author, E-Mail: joshi@iitk.ac.in.

## Abstract

In this paper, ageing behavior of suspensions of laponite with varying salt concentration is investigated using rheological tools. It is observed that the ageing is accompanied by an increase in the complex viscosity. The succeeding creep experiments performed at various ages showed damped oscillations in the strain. The characteristic time-scale of the damped oscillations, retardation time, showed a prominent decrease with the age of the system. However, this dependence weakens with an increase in the salt concentration, which is known to change microstructure of the system from glass-like to gel-like. We postulate that a decrease in the retardation time can be represented as a decrease in the viscosity (friction) of the dissipative environment surrounding the arrested entities that opposes elastic deformation of the system. We believe that ageing in colloidal glass leads to a greater ordering that enhances relative spacing between the constituents thereby reducing the frictional resistance. However, since a gel state is inherently different in structure (fractal network) than that of a glass (disordered), ageing in the same does not induce ordering. Consequently, we observe inverse dependence of retardation time on age becoming weaker with an increase in the salt concentration. We analyze these results from a perspective of ageing dynamics of both glass state and gel state of laponite suspensions.

## I. INTRODUCTION

Glassy state is generally referred to a state wherein the system explores only a part of the phase space available to it (Liu & Nagel 1998). Despite many decades of active research aimed at improving our understanding of these systems, various intriguing features of the glassy state dynamics are still poorly understood and thus continue to attract significant attention from the physics, the biology and the engineering community (Cipelletti & Ramos 2005; Sciortino & Tartaglia 2005). Apart from molecular glass formers, non-ergodic systems that dominate the present research activity in this field include colloidal suspensions (Weeks et al. 2007), polymer/clay composites (Treece & Oberhauser 2007), emulsions (Gang et al. 1999), gels (Cloitre et al. 2003), foams (Cohen-Addad & Hohler 2001), block copolymers (Mallamace et al. 2000), etc. Furthermore, in addition to the significant academic interest in understanding the glassy-state dynamics in these systems, their industrial applications have provided further impetus for such studies.

In this work, we investigate the ageing dynamics of aqueous laponite suspensions using rheological tools. Laponite is composed of disc shaped nanoparticles with a diameter 25 nm and layer thickness 1 nm (Kroon et al. 1998). The chemical formula for laponite is $Na^+_{0.7}[(Si_8Mg_{5.5}Li_{0.3})O_{20}(OH)_4]^-_{0.7}$. An isomorphic substitution of magnesium by lithium atoms generates negative charge on its surface that is counterbalanced by the positive charge of the sodium ions present in the interlayer (Van Olphen 1977). In an aqueous medium, sodium ions dissociate which, in turn, leads to a net negative charge on its surface. The edge of the laponite particle is composed of hydrous oxide and its charge is less negative in the basic pH medium while positive in the acidic pH medium (Van Olphen 1977). Its suspension in aqueous medium leads to the formation of non-ergodic soft solids which show a rich variety of physical behavior.

Aqueous laponite suspensions display a very rich phase behavior and different groups have proposed various versions of phase diagrams for laponite with respect to the concentrations of the salt and of the laponite (Gabriel et al. 1996; Michot et al. 2006; Mongondry et al. 2005; Mourchid et al. 1995a; Ruzicka et al. 2006; Ruzicka et al. 2007; Tanaka et al. 2004). In the salt free aqueous medium and at pH 10 (~$10^{-4}$ M $Na^+$ ions), the electrostatic screening length associated with a laponite particle is around 30 nm (Bonn et al. 1999). Under such conditions the net interaction between various laponite particles is repulsive in nature. It is generally believed that for concentrations above 1 wt. % and in the near absence of any salt in the system (or below $10^{-4}$ M concentration of Na$^+$ ions), ergodicity breaking leads to the formation of the repulsive (Wigner) glasses (Bonn et al. 1998; Knaebel et al. 2000; Levitz et al. 2000; Tanaka et al. 2004). On the other hand, the addition of a salt to a laponite suspension increases the concentration of cations in the system which screen the negative charges on the laponite particle thereby changing the state of the system from being a repulsive glass to a gel (Avery & Ramsay 1986; Kroon et al. 1998; Kroon et al. 1996; Mongondry et al. 2005; Mourchid et al. 1995a; Mourchid et al. 1995b; Nicolai & Cocard 2000; Nicolai & Cocard 2001; Pignon et al. 1997a; Pignon et al. 1997b; Pignon et al. 1996). At higher concentration of laponite, system enters into a nematic phase (Bhatia et al. 2003; Gabriel et al. 1996; Ravi Kumar et al. 2007). Generally a glassy state is distinguished from a gel state based on the presence of fractal network in the latter, while in the case of former density is uniform for probed length scales greater than the particle length scale (including the Debye screening length).

During the past few years, several groups have studied the ergodic-nonergodic transition of a laponite suspension at low ionic concentrations using various optical and rheological techniques (Mongondry et al. 2005; Ruzicka et al. 2006; Schosseler et al. 2006; Tanaka et al. 2005). A common procedure is to pass the laponite dispersions through micro-filters to study the structural evolution with respect to its "age." It is generally acknowledged that the filtration influences their optical characteristics but not their rheological response (Bonn et al. 1999). It is observed that the system has two relaxation modes. The fast or $\beta$ - mode is observed to be independent of the age of the sample (Abou et al. 2001; Bellour et al. 2003); however, Ruzicka and coworkers (Ruzicka et al. 2004) have recently reported a slight increase in this mode with age in the low concentration regime (<1.5 – 1.8 wt. %). The slow or $\alpha$ - mode of relaxation shows an initial rapid increase followed by a linear increase with age (Abou et al. 2001; Bellour et al. 2003; Schosseler et al. 2006; Tanaka et al. 2005). The sub-regime with a rapid initial increase is called a cage forming regime while the latter regime is called the full aging regime (Tanaka et al. 2005). Recently, Joshi (Joshi 2007) proposed that the rapid increase in the relaxation time with age in the cage formation regime is due to the osmotic swelling of laponite clusters that are formed immediately after the preparation of the sample.

Owing to the widespread applications of aqueous clay suspensions, significant work has been carried out on their rheological characterization. Most of the early studies concentrated on thixotropy, yield stress and pumpability of these materials. van Olephan (Van Olphen 1977) and recently Barnes (Barnes 1997) and Luckham and Rossi (Luckham & Rossi 1999) have all presented excellent state of the art accounts of most of the previous rheological works available in the literature. Very recently Coussot (Coussot 2006; Coussot 2007) has provided an excellent summary of recent literature on rheological behavior and modeling approaches of ageing dynamics of soft non-ergodic materials including clay suspensions. Lately focus of various studies is dominated by understanding of their micro-structural description through rheological measurements. Cocard et al. (Cocard et al. 2000) studied frequency dependence of time evolution of elastic modulus for ageing laponite suspension which is in a gel state at various salt concentrations. They also observed that frequency dependence of elastic and viscous modulus, which is similar to that of viscous liquid at small age, weakens with age (or gelation time). They observed that the rate of gelation increases with ionic strength. Bonn *et al.* (Bonn et al. 2002b) studied aqueous laponite suspensions in a basic medium using in situ diffusive wave spectroscopy. They observed that the viscosity dependence of age directly correlates with the slow mode relaxation time dependence of age. Subsequently, in a similar observation to that of Cocard et al.(Cocard et al. 2000), they (Bonn et al. 2002a) reported that immediately after sample preparation, the frequency dependence of the viscous and elastic moduli is similar to that of a viscous liquid. Eventually both moduli become independent of the frequency as the glass transition is approached. Mourchid *et al.* (Mourchid et al. 1995a) used the shear rheology to monitor sol-gel transition in laponite suspensions and observed that increasing the ionic strength shifts the sol-gel transition to a lower volume fraction.

In creep experiments, many soft glassy systems show damped oscillations in strain due to a coupling between the viscoelastic behavior of the fluid and the inertia of the instrument. Baravian and Quemada (Baravian & Quemada 1998) systematically analyzed such inertial oscillations using various linear viscoelastic mechanical models. Subsequently, Baravian *et al.* (Baravian et al. 2003) studied the creep behavior of aqueous montmorilloinite suspensions. By analyzing the inertial oscillations, they estimated high frequency elastic modulus of the system, which otherwise was not possible using the conventional oscillatory tests. Recently Coussot *et al.* (Coussot et al. 2006) studied three complex fluids, namely an aqueous bentonite suspension, mustard and hair gel using creep experiments and observed oscillations in strain that attenuate very fast. They estimated the elastic modulus of the system which shows logarithmic dependence on age. In a subsequent work, they (Ovarlez & Coussot 2007) studied the effect of temperature, density, and concentration on the ageing behavior of aqueous bentonite suspension. They observe that all the elastic modulus versus time curves fall on a single master curve when rescaled by a factor which is function of concentration, temperature, strength of deformation and age of the system, demonstrating equivalence of these parameters in the context of ageing behavior. Recently Joshi and Reddy (Joshi & Reddy 2007) carried out systematic creep experiments on ageing soft solids of laponite at various ages and stresses. They observed that when imposed time is normalized by dominant relaxation mode of the system that depends on age and stess, universal master curve for creep is obtained which is invariant of applied stress and age of the system. Similarly, in recent years considerable work has been carried out to understand the rejuvenation phenomena *per se* in soft glassy systems (Abou et al. 2003; Bonn et al. 2002b; Cloitre et al. 2000; Di Leonardo et al. 2005; Ianni et al. 2007; Viasnoff & Lequeux 2002). However, various features of such ageing systems such as over-ageing, (Viasnoff & Lequeux 2002) multiple decoupled timescales and their distribution (Abou et al. 2001), complicated energy landscapes leading to multiple ageing paths (Jabbari-Farouji et al. 2007), fast aging dynamics of rejuvenated systems (Ianni et al. 2007), inapplicability of time translational invariance (Fielding et al. 2000), etc. demand further study and understanding of the ageing phenomena.

In this work, we perform oscillatory and creep experiments on laponite suspensions to investigate their ageing dynamics after carrying out a specific rejuvenation procedure. In particular, we exploit the information embedded in the damped oscillations of strain in creep to estimate the retardation time of the system. A retardation time represents resistance offered by the microstructure to the elastic deformation of the system. As the retardation time tends to zero, thereby meaning no resistance, the system response becomes perfectly elastic. We observe that as a system ages, although the resistance to translational diffusion significantly increases, at a local level the friction decreases leading to a lowering of the retardation time with age. Furthermore, we also analyze the dependence of the retardation time on the age of the system and on the molar concentration of salt that reveals distinguishable characteristics of ageing in glasses and gels. We believe that this information provides further useful insights into the underlying microscopic phenomena in the process of ageing in such systems.

## II. PREPARATION AND VISCOMETRY

Laponite RD, synthetic hectorite clay, used in this study was procured from Southern Clay Products, Inc. A predetermined specific molar concentration of $Na^+$ ions was maintained by adding NaCl to ultra pure water. The white powder of Laponite was dried for 4 hours at 120 °C before mixing it with water at pH 10 under vigorous stirring conditions. The basic pH (~10) was maintained by the addition of NaOH to provide chemical stability to the suspension. The suspension was stirred vigorously for 15 min. We have used seven systems, six systems have 3.5 wt. % laponite and $10^{-4}$, $10^{-3}$, $3\times10^{-3}$, $5\times10^{-3}$ M, $7\times10^{-3}$ M and $10^{-2}$ M concentration of $Na^+$ ions while the seventh system has 2 wt. % laponite and $10^{-2}$ M ionic concentration of the $Na^+$ ions.

In this work, stress controlled oscillatory shear experiments and creep experiments were carried out using a stress controlled rheometer, AR 1000 (Couette geometry, bob diameter 28 mm with gap 1mm). The couette cell was filled up with the test sample and 3.5 %, $10^{-4}$ M and 2.0 %, $10^{-2}$ M samples were left to age for 3 hours, while 3.5 %; $10^{-3}$ M, $3\times10^{-3}$ M, $5\times10^{-3}$ M, $7\times10^{-3}$ M and $10^{-2}$ M samples were left to age for 90 mins. To avoid the loss of water by evaporation or the possibility of $CO_2$ contamination of the sample, the free surface of the suspension was covered with a thin layer of a low viscosity silicon oil during the course of viscometric measurements. Subsequently, we applied an oscillatory deformation with stress amplitude of 50 Pa and frequency 0.1 Hz for about 600 s to 1100 s unless otherwise mentioned. As expected, the suspension yields under such a high stress and eventually shows a plateau of low viscosity that does not change with time. We stopped the rejuvenation (shear melting) experiment at this point in time, from which the aging time was measured. Subsequent to rejuvenation, we carried out the stress controlled oscillatory shear experiments by employing a shear stress amplitude of 0.5 Pa and frequency 0.1 Hz to record the ageing behavior of this system. In all creep experiments, we have applied a constant stress of 1.5 Pa for 100 s unless otherwise mentioned. All results reported in this paper relate to 20 °C. We also carried out additional frequency sweep experiments on independent samples at the end of the waiting period. It was observed that the storage modulus was independent of the frequency while the loss modulus showed a slight decrease with frequency in the experimentally accessible frequency range. This observation is in line with the findings of Bonn *et al.* (Bonn et al. 2002a), and according to Fielding *et al.* (Fielding et al. 2000) this ensures system to be in the non-ergodic regime.

**III. RESULTS AND DISCUSSION**

In this work, we have employed a specific experimental protocol to study the ageing behavior of a laponite suspension. Freshly prepared laponite samples were left idle for a fixed period of time as mentioned in the previous section in a couette geometry, prior to being subjected to a rejuvenation (shear melting) experiment by applying an oscillatory stress of 50 Pa. Figure 1 shows a typical shear rejuvenation behaviour for several independent (freshly prepared) samples of 3.5 %, $10^{-3}$ M laponite suspension. It can be seen that initially the viscosity gradually decreases, followed by a very sharp decrease until a plateau is reached that does not change with time. For this system, we have varied the rejuvenation time from 600 s to 900 s; however, each experiment was stopped only after the plateau was reached. Inset in figure 1 shows the corresponding ageing curves, where the age is measured after rejuvenation experiment is stopped. It can be seen that the ageing behavior is relatively insensitive to the minor variations in the rejuvenation step. In the shear rejuvenation experiments, the primary harmonic in strain was at least an order of magnitude larger than the third harmonic and hence it is justified to analyze these data in terms of a complex viscosity. Figure 2 shows a typical shear rejuvenation and the subsequent ageing behaviour for several independent samples of 3.5 %, $5\times10^{-3}$ M laponite suspension. Since the stress of 50 Pa was not sufficient to cause yielding in this system, an oscillatory stress with the magnitude of 70 Pa was employed. Interestingly, this system shows yielding in two stages, with intermediate plateau occurring at a complex viscosity of around 5 Pas. For one sample (shown in filled diamonds), oscillatory stress of 70 Pa was applied for 178 s followed by 80 Pa for 307 s followed by 70 Pa again for 386 s, in order to avoid the formation of an intermediate plateau. The data for which 80 Pa stress was applied is shown by gray shade in the same figure. Subsequent ageing curve is shown in an inset where oscillatory stress of 0.5 Pa was employed. It can be seen that the various ageing curves are very similar irrespective of the differences inherent in the rejuvenation step. For the 3.5 %, $7\times10^{-3}$ M and $10^{-2}$ M laponite suspensions, we used different stress amplitudes to rejuvenate the samples, however observed no difference in the ageing behavior. For the 2 % system at $10^{-2}$ M ionic concentration, we have employed the stress amplitude of 40 Pa in the rejuvenation step.

An important feature of the rejuvenation step can be summarized as follows: once a plateau of low viscosity is reached (which does not change with time), the ageing behavior of the system becomes independent of the minor differences and the time elapsed during the rejuvenation experiment. In the ageing system, various properties of the system are known to evolve with its age. In general, it is observed that the characteristic (dominant) relaxation time of the system changes with age as: $\tau_\alpha \sim t_w^\mu$, where $t_w$ is age and $\mu$ is a positive constant (Cloitre et al. 2000; Fielding et al. 2000; Struik 1978). Struik (Struik 1978) suggested that in the limit $\mu \to 1$, simple ageing occurs while in the limit $\mu \to 0$, obviously, no effect of ageing is seen (also see Fielding et al. (Fielding et al. 2000)). For a soft micro-gel paste, Cloitre et al. (Cloitre et al. 2000) observed that when a vanishingly small stress is applied to the system, one obtains the limit $\mu = 1$, while in the limit of very large stresses, one recovers the system that does not age ($\mu = 0$) until the stress is removed. Furthermore, a strong stress field leading to plastic deformation, as in the present case, is known to erase the deformation history of the sample (Utz et al. 2000), and also, it does not let the system age (as plateau viscosity does not change with time). Thus, the present observation that the minor differences in rejuvenation step do not affect the long term ageing behavior (as long as a strong stress field is applied in the rejuvenation) is not surprising. Interestingly, Cloitre et al. (Cloitre et al. 2000) have also reported similar observations for a microgel paste.

Figure 3 shows the evolution of the complex viscosity for a shear rejuvenated laponite suspension for a range of molar concentrations of NaCl. It is evident that the systems with higher salt concentration are stiffer when compared at same age. Furthermore, the complex viscosity shows a power-law type dependence on age ($\eta^* \sim t_w^x$). The corresponding power law exponent ($x$) is plotted against the molar concentration of $Na^+$ ions as an inset in the same figure. It can be seen that the complex viscosity shows a weaker dependence on age with the increase in the salt concentration. Pignon et al. (Pignon et al. 1997b) also observed that storage and loss modulus of systems with higher salt concentration is higher at the same age, however they did not report the rate of increase of the same with age. We also carried out experiments with 2 % laponite suspensions with $10^{-2}$ M ionic concentration, whose ageing behavior is represented by open circles in figure 3. The latter sample shows a different ageing behavior compared to the 3.5 % samples. It should be noted that at sufficiently higher age, the contribution of the viscous modulus to the complex viscosity is negligible, which makes $G'$ vary with age in a similar fashion as the complex viscosity as shown in figure 3.

In a non-ergodic state, owing to their disc- like shape and non- uniform charge distribution, laponite particles can be considered to be trapped in a cage of surrounding particles, which can be represented by a potential energy well. Distribution of such local energy minima is known as an energy landscape (Fielding et al. 2000). Although there is no global minimum, each laponite particle undergoes an activated dynamics of structural rearrangements so that the system attains a lower energy state with age making the system more elastic. An increase in salt concentration decreases the electrostatic screening length associated with the laponite discs, changing the energy landscape and its evolution with age. It is generally believed that an increase in the salt concentration leads to the formation of a gel-like state, while for the system with no salt present, a glass-like state is formed. The ageing dynamics shown in figure 3 essentially captures this dynamics; wherein an increase in salt concentration changes the evolution of complex viscosity (elastic modulus) with age. Cocard et al. (Cocard et al. 2000) reported that the rate of gelation increases with ionic strength. However, if the increase in complex viscosity with age is considered as a signature of ageing (or gelation at higher ionic strength), we observe that after sufficient age, rate of gelation gets slower with increase in ionic strength.

We carried out aging experiments for all samples until the complex viscosity reached a predetermined value, typically between 300 Pas to 1750 Pas. After stopping the oscillatory test each time, creep experiment was performed. Figure 4 shows a typical creep curve for a sample having a complex viscosity of 300 Pas. In the initial period up to $O(1)$ s, the system shows significant oscillations in strain which attenuate very quickly. Similar oscillations are also observed in the recovery experiments. Such behavior is known to occur as a result of visco-elastic character of the fluid coupled with the instrument inertia. Baravian and Quemada (Baravian & Quemada 1998) have analyzed this so-called creep ringing behavior for various linear viscoelastic models. The essential features of this approach are recapitulated here. The application of the Newton's second law of motion to a rheometer system leads to (Baravian & Quemada 1998):

$$a\ddot{\gamma} = \sigma_0 - \sigma, \qquad (1)$$

where $a$ is the moment of inertia of the mobile part, $\gamma$ is shear strain while $\sigma_0$ and $\sigma$ are the stresses corresponding to the applied and effective torques respectively. The Maxwell-Jeffreys constitutive relation (mechanical model shown in figure 5), written as,

$$(\eta_1 + \eta_2)\dot{\sigma} + G\sigma = \eta_1\eta_2\ddot{\gamma} + \eta_2 G\dot{\gamma}, \qquad (2)$$

coupled with eq. 1 can be solved analytically for the creep flow with a constant shear stress $\sigma_0$. The resultant shear strain $\gamma$ is given by (Baravian & Quemada 1998):

$$\gamma(t) = \sigma_0 \left\{ \frac{t}{\eta_2} - B + e^{-At} \left[ B\cos\bar{\omega}t + \frac{A}{\bar{\omega}} \left( B - \frac{1}{A\eta_2} \right) \sin\bar{\omega}t \right] \right\}, \qquad (3)$$

where $A = \dfrac{aG + \eta_1\eta_2}{2a(\eta_1 + \eta_2)}$, $B = \dfrac{a(\eta_1 + \eta_2)}{\eta_2 G}\left(\dfrac{2A}{\eta_2} - \dfrac{1}{a}\right)$ and $\bar{\omega} = \sqrt{\dfrac{\eta_2 G}{a(\eta_1 + \eta_2)} - A^2}$. It can be seen that for real values of $\bar{\omega}$, shear strain undergoes damped oscillations in a creep experiment as shown in figure 4. Furthermore, damped oscillatory response attenuates quickly, which is followed by a usual creep behavior.

Inset in figure 4 shows a fit of eq. 3 to the damped oscillatory strain response plotted in figure 3 in the initial stages (up to 0.5 s). It can be seen that eq. 3 provides an excellent fit to the oscillatory data, from which model parameters $\eta_1$, $\eta_2$ and $G$ can be obtained at various ages of the sample. The value of $a$ is obtained by measuring the strain for the applied stress without any sample as suggested by Coussot et al. (Coussot et al. 2006). Furthermore, similar to the findings of Coussot *et al.* (Coussot et al. 2006), the values of $G$ and $G'$ were seen to be very similar. Eq. 3 was fitted to the experimental data by using Levenberg-Marquardt method for nonlinear least squares fitting of the commercial software Origin® 7. It should be mentioned here that the fit was not particularly sensitive to the value of $\eta_2$ over several decades. This is not at all surprising since the experimental data used in the fitting is limited only up to 0.5 s, over which the deformation is predominantly elastic in nature. In order to get the exact value of $\eta_2$, we made use of the data obtained from the oscillatory experiments. Findley *et al.* (Findley et al. 1976) have discussed the oscillatory response of the Maxwell-Jeffreys model. If $p = (\eta_1 + \eta_2)/G$ and $q = \eta_1\eta_2/G$, then the storage and loss moduli of the Maxwell-Jeffreys model are given by:

$$G' = \frac{p\eta_2\omega^2 - q\omega^2}{p^2\omega^2 + 1}, \quad G'' = \frac{pq\omega^3 + \eta_2\omega}{p^2\omega^2 + 1}. \qquad (4)$$

For the range of values of $\eta_2$ for which an excellent fit to the oscillatory behavior is obtained, we observed that: $\eta_1 \ll \eta_2$, $G^2/\eta_2 \gg \eta_1\omega^2$, $a/\eta_2^2 \ll 1/G$ and $G/\eta_2 \ll \eta_1/a$. This in turn leads to considerable simplifications of the above expressions of the elastic and viscous modulii as follows:

$$G' \approx G, \quad G'' \approx \frac{G^2}{\eta_2\omega}. \qquad (5)$$

Thus, the values of $\eta_2$ as a function of age can be easily obtained from the oscillatory data via eq. (5). Thus following above mentioned procedure all the model parameters of Maxwell-Jeffreys model can be estimated with respect to age. It should be noted that we carry out the oscillatory experiments at frequency 0.1 Hz. If the experiments would have been performed at very high frequency such that $G^2/\eta_2 \ll \eta_1\omega^2$, we would have recovered Kelvin Voigt response (Maxwell-Jeffrey model minus dashpot) from the oscillatory behavior. However, performing experiments at high frequencies may not be always possible with rheometer due to instabilities that set in due to inertia effects (Marin 1988). Kelvin-Voigt response can also be recovered in the limit of $\eta_2 \to \infty$. However in the present case we get response expressed by equation (5), which is different from Kelvin-Voigt response, as $\eta_2$ is not large enough and time scale of deformation is comparatively large (that is $\omega$ is small). Thus the present experimental scheme gives us an advantage in order to estimate $\eta_2$ from oscillatory experiments, which otherwise is very difficult to measure due to scatter in steady shear data at low stresses and rejuvenation at high stresses.

Figure 6 shows $\eta_2$ for various salt concentrations as a function of age. Included in this figure are also the results for the system without any salt. It can be seen that $\eta_2$ increases more rapidly with age and it is significantly larger in magnitude than the complex viscosity. Inset of figure 6 shows power law exponent for the dependence of $\eta_2$ on age plotted against molar concentration of Na$^+$ ions. It can be seen that the dependence of $\eta_2$ on age becomes weaker with the increasing salt concentration; this trend is similar to that of the dependence of the complex viscosity on age. However, for the molar concentration 10$^{-2}$ M, the exponent shows a sudden increase. This may be due to the fact that the 3.5 % system flocculates at about this molar concentration of the salt (Mongondry et al. 2005; Mourchid et al. 1995a; Ruzicka et al. 2006; Tanaka et al. 2004).

The Maxwell-Jeffreys constitutive relation has two time scales associated with it. The slow time scale is associated with the shear viscosity $\eta_2$ and shear modulus $G$, while the second time scale, also known as the retardation time scale is linked with the viscosity $\eta_1$ and shear modulus $G$. The former time scale is given by:

$$\tau_\eta = \eta_2/G. \tag{6}$$

Figure 7 shows the dependence of this time scale on age for a range of values of the concentration of salt. It is readily seen that the value of $\tau_\eta$ is significantly smaller than the age of the system. In the literature, it is believed that $\tau_\eta$ has same magnitude as the α relaxation mode of the system. However, the present results are at variance from this trend, as not only its magnitude is significantly smaller than its age, but its dependence on age is also weaker than the linear dependence reported in the literature.

The second characteristic time of the system, retardation time, can be obtained from $\eta_1$ and $G$, and is defined as (Barnes et al. 1989; Harrison 1976):

$$\tau_r = \eta_1/G. \tag{7}$$

The retardation time controls the rate of growth of strain following the imposition of the stress (Barnes et al. 1989). As the Maxwell-Jeffreys model depicted in figure 5 suggests, it is essentially related to the viscous environment (or friction) associated with the elastic component of the system. As discussed before, in the non-ergodic state, individual charged laponite particles can be considered to be trapped in a potential energy well of surrounding particles. Such entrapment of particles is essentially responsible for the elasticity of such an ageing suspension. When a step stress is applied to the system, the particles change their position by working against the potential energy field. The dissipative environment surrounding the particles resists any sudden change in the position and retards this process. Thus, the retardation time denotes the contribution of friction while the system is undergoing elastic deformation. More is the friction that particles encounter upon the application of stress, greater is the retardation time. A spectrum of retardation times is represented by a series combination of several Kelvin-Voigt elements (Maxwell-Jeffreys mechanical model minus the dashpot) (Barnes et al. 1989; Harrison 1976). However, for the present system, only a single Kelvin-Voigt element fits the oscillatory data very well thereby suggesting that the retardation time of the system is nearly monodispersed.

Figure 8 shows the dependence of $\tau_r$ on age for the 3.5 %, $10^{-4}$ M sample for various creep stress levels, namely, 0.5, 1.5 and 3 Pa. It can be seen that retardation time decreases with respect to age. This result is particularly significant as it suggests a reduction in friction in the neighborhood of arrested particles showing a decrease in resistance to elastic deformation with age. Furthermore, the dependence of the retardation time on age ($\tau_r \sim t_w^{-0.6}$) is independent of the applied stress. We also confirm this result by carrying out additional creep tests on suspensions having different molarity at various stresses. We believe that, since the response time of the damped oscillations is $O(1 \text{ s})$, stress may not be able to influence the fluid behavior, as deformation is essentially elastic over this time scale, thereby showing the retardation time to be independent of the applied stress.

An evolution of the energy landscape giving more weightage to the lower energy states with age is responsible for an increase in the elasticity of the system as shown in figure 3. This inherent elastic character of the system coupled with the inertia of the system is responsible for an oscillatory response to the applied step stress which is damped by the dissipative environment (friction) surrounding the particles leading to the overall damped response. Figure 9 shows the dependence of $\tau_r$ on age for a range of salt concentrations. In figure 10, power law exponent $b$ for $\left(\tau_r \sim t_w^b\right)$ is plotted with respect to the ionic concentration. Both these figures show that $\tau_r$ significantly decreases with age at low salt concentrations. Similar to the trend observed for $\eta_2$, power law exponent for ionic concentration $10^{-2}$ M shows a sudden change in the behavior which may be due to flocculation.

Inset of figure 9 shows $\tau_r$ plotted against complex viscosity for two systems having 3.5 % laponite along with a system having 2 % laponite and molarity $10^{-2}$ M in order to get a feel of the dependence and magnitudes of the retardation time on concentration at the same level of complex viscosity. The values of $\tau_r$ for the 3.5 % suspensions at other salt concentrations were found to be of the same order of magnitude as the two 3.5 % systems shown in the inset of figure 9; hence, these data are not included here to avoid crowding of the figure. It can be seen that the value of $\tau_r$ for the 2 % system is significantly smaller than that of the 3.5 % system. This suggests that the value of $\eta_1$ of the 2 % system must also be smaller than that of the 3.5 % systems at a constant value of $G$ (or $\eta^*$). This implies that a decrease in laponite concentration causes a decrease in $\eta_1$. This observation further strengthens our proposal that $\eta_1$ may be associated with dissipative environment or friction and increases with an increase in the laponite concentration.

Light scattering studies on ageing laponite suspensions generally characterize fast and slow time scale dependence on age (Abou et al. 2001; Bellour et al. 2003; Kaloun et al. 2005; Schosseler et al. 2006; Tanaka et al. 2005). In light scattering experiments autocorrelation function shows two stage decay. Normalized autocorrelation function is then fitted by an empirical equation having a sum of an exponential and a stretched exponential function to yield *α* and *β* relaxation timescales (Abou et al. 2001). The fast or *β* time scale is considered to be related to the rattling motion of the trapped entity and is observed to be independent of age (Schosseler et al. 2006; Tanaka et al. 2005). It is however worthwhile to add here that the retardation time $\tau_r$ is not same as $\beta$ mode, as the latter remains constant with age. On the other hand, the slow or $\alpha$ time-scale represents the time-scale of cage diffusion and is observed to increase linearly with age under static conditions (Knaebel et al. 2000; Schosseler et al. 2006). It is interesting to note that, although the $\alpha$ time scale increases with age due to the deepening of the energy well, the retardation time actually decreases suggesting a decrease in the resistance to elastic deformation. The activated dynamics that jammed particles undergo might be responsible for positioning them in the cooperative environment of cage such that resistance to elastic deformation decreases indicating a decrease in local friction surrounding the particles. Thus, the retardation time as used in the present context provides an additional vantage point to study the phenomenon of ageing in such systems.

In order to gain further insights about the dependence of $\eta_1$, we have plotted $\eta_1$ with age for various salt concentrations in figure 11. Figure 10 shows the corresponding dependence of power law exponent $a$ for $\left(\eta_1 \sim t_w^a\right)$ with respect to the ionic concentration. In general, it can be seen that the dependence of $\eta_1$ on age shows a similar trend to that of $\tau_r$. It should be noted that, while discussing retardation behavior, we are concentrating on creep time up to $O$ (1s) thereby focusing only on the elastic deformation of an ageing laponite suspension. The frictional contribution, represented by a dashpot with viscosity $\eta_1$ in figure 5, contributes towards resisting elastic deformation of the system. Figure 10 shows that $\eta_1$ is a decreasing function of age up to molar concentration of Na$^+$ ions of 3×10$^{-3}$ M. Beyond that this dependence weakens significantly. A decrease in $\eta_1$ with age suggests a reduction in frictional contribution that resists elastic deformation of the system. We believe that this frictional contribution represented by $\eta_1$ arises from the effective viscosity of the medium surrounding the laponite particles, which seem to be decreasing with age as laponite particles undergo activated dynamics and progressively attains lower energy. However, this frictional contribution is different from the overall resistance the system offers to deformation at larger time scale represented by $\eta_2$, which is observed to be strongly increasing with age as shown in figure 6. This result is particularly very significant since ageing is known to enhance resistance to translational diffusion, however present results suggests that this is accompanied by a reduction in friction in the neighborhood of the arrested entity.

It is generally believed that in the absence of any salt, repulsive interactions prevail among laponite particles that lead to the formation of colloidal Wigner glass (Mourchid et al. 1995a; Mourchid et al. 1995b; Schosseler et al. 2006; Tanaka et al. 2004). In the glassy state, the system is in nonergodic regime due to structural arrest caused by strong repulsive interactions. As the concentration of $Na^+$ ions increases, it gradually screens electrostatic repulsion and at sufficiently high concentration of $Na^+$ ions, attractive interactions prevail. This causes fractal network of laponite particles that spans the whole space leading to a gel state (Nicolai & Cocard 2001; Pignon et al. 1997a; Pignon et al. 1997b; Pignon et al. 1996). Figure 12 shows schematic representation of a glassy state and a gel state, as postulated by Tanaka et al. (Tanaka et al. 2004). The critical salt concentration delineating the boundary between the glass state and the gel state is however far from obvious and as such is a matter of debate (Mongondry et al. 2005; Mourchid et al. 1995a; Ruzicka et al. 2006; Ruzicka et al. 2007; Tanaka et al. 2004). Accordingly it is obvious that the presence of $Na^+$ ions alters the energy landscape significantly and as a consequence the activated dynamics that causes ageing in the nonergodic state.

In a light scattering study of laponite suspension, Nicolai and Cocard (Nicolai & Cocard 2001) observed that for ionic concentrations above $10^{-3}$ M, characteristic length scale of the network (gel), correlation length, increases with age. They further observed that correlation length increases with ionic strength while decreases with concentration of laponite. However at lower ionic concentration, static structure factor was observed to decrease with age indicating an increase in the repulsive interactions (Tanaka et al. 2004). Very recent study of Brownian dynamics simulations by Mossa et al., (Mossa et al. 2007) suggested that ageing dynamics strongly affects orientational degree of freedom which relaxes over the timescale of translational modes. The rheological study described in this work gives further insights into the ageing dynamics. We observe that as repulsive interactions decrease gradually (with an increase in salt concentration), the rate of decrease of the retardation time with age, which signifies resistance to elastic deformation or friction localized in the neighborhood of the particle, becomes weak. Furthermore, not just retardation time but all the other characteristics of the system including complex viscosity, show weaker evolution with respect to age with an increase in the salt concentration. This observation indicates that activated dynamics is faster in a glass state than in a gel state.

The present work clearly distinguishes between ageing in a glass and a gel. In structural glasses ageing leads to densification by causing a greater order in the molecular arrangement. In colloidal glasses, although density remains constant, ageing is expected to cause progressive ordering which should lead to more relative space between the constituents that may cause a reduction in resistance to elastic deformation. We propose that as colloidal glass ages it leads to a more ordered state with respect to age, though complete crystallization may not be possible and thus demonstrates the observed behavior. However as we increase the salt concentration, state of the suspension changes to a gel. Since a gel state is inherently different in structure (fractal network) than that of a glass (disordered), ageing in the same does not lead to ordering and we observe a weaker dependence of retardation time on age with increase in the salt concentration. Therefore the present work differentiates between a glass and a gel by their ageing behavior such that a glass forms greater ordered structure upon ageing which is not likely in the case of a gel. This is a very significant observation which elucidates an important difference between a glassy state compared to a gel state, a matter that has been debated in the literature at a great length. We believe that various results described in the present work provide significant additional insight into the ageing dynamics of colloidal glasses and gels that may provide stimuli for further studies on the soft glassy materials.

## IV. SUMMARY

Ageing dynamics of aqueous laponite suspensions at various salt concentrations is studied using rheological tools by employing a well-defined rejuvenation procedure. It was observed that for large values of the rejuvenating stress, minor differences in the rejuvenation procedure do not affect the ageing behavior of laponite suspensions. Subsequent ageing experiments were carried out until a predetermined complex viscosity value is reached and then creep experiments were performed. In general we observe that shear viscosity is significantly higher in magnitude and increases very rapidly than the complex viscosity. In creep experiments, soft solids of laponite show oscillations in strain that attenuate over a time period $O(1 \text{ s})$ due to the coupling of the viscoelasticity of the sample with the instrument inertia. A single mode Maxwell-Jeffreys model coupled with inertia gives an excellent fit to the damped oscillation. The characteristic time scale of the Kelvin-Voigt element of the Maxwell-Jeffreys model, also known as the retardation time, ($\tau_r = \eta_1/G$) is seen to be independent of the applied stress; however, it decreases with the age of the sample. Furthermore, the rate of decrease of $\tau_r$ became weaker with an increase in salt concentration. The 2 % laponite showed significantly lower values of $\tau_r$ compared to 3.5 % suspensions. In the present context, the retardation time is associated with the time required for a transient to attenuate while approaching a steady state, and it represents a frictional contribution of the viscous medium surrounding the particles that resists elastic deformation of the system. Notably, these results show that when the resistance to translational diffusion of the arrested entities increases with age, at local level, the friction actually decreases. This decrease becomes weaker with an increase in the salt concentration which is known to change the system from a glass state to a gel state. In addition, we observe that the ageing dynamics in the glassy state is faster than that in the gel state. Simple analogy of colloidal glasses with molecular glasses suggests that ageing should induce greater ordering which provides more space for the constituents of the glass, resulting in a lesser resistance to the elastic deformation. However as we increase the salt concentration, state of the suspension changes to a gel. Since a gel is comprised of a fractal network, ageing in the same does not lead to an ordered structure and we observe a weaker dependence of retardation time on age with increase in the salt concentration. This study clearly shows that the use of rheological tools can certainly provide useful insights thereby leading to a better understanding of the ageing behavior in soft glassy materials.

**Acknowledgement**: This work was supported by the IIT Kanpur young faculty initiation grant and BRNS young scientist research project awarded by Department of Atomic Energy, Government of India to YMJ.

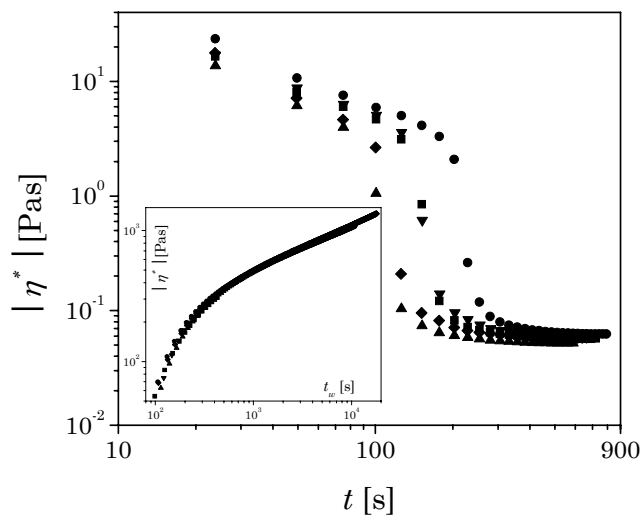

**Figure 1.** Rejuvenation data for 3.5 %, $10^{-3}$ M laponite suspension. Rejuvenation was carried out at stress amplitude 50 Pa. Inset shows corresponding ageing curves at stress amplitude 0.5 Pa that is carried out subsequent to rejuvenation experiment. Same symbol is used for rejuvenation and ageing of the same sample.

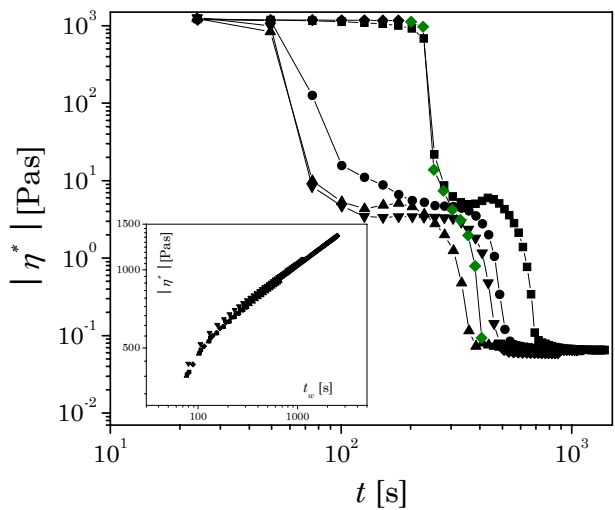

**Figure 2.** Rejuvenation data for 3.5 %, $5\times 10^{-3}$ M laponite suspension. Rejuvenation was carried out at stress amplitude 70 Pa. For filled diamonds, initially stress amplitude of 70 Pa was used. It was changed to 80 Pa for a brief period of time and changed back to 70 Pa again (refer to text for details). The points for which 80 Pa stress amplitude was employed is shown in gray shade. Line is a guide to an eye. Inset shows corresponding ageing curves at stress amplitude 0.5 Pa that is carried out subsequent to rejuvenation experiment. Same symbol is used for rejuvenation and ageing of the same sample.

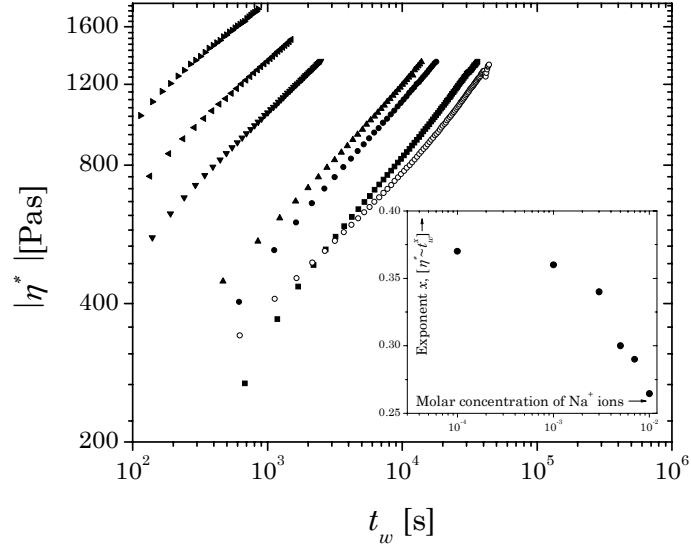

**Figure 3.** Evolution of complex viscosity with age after rejuvenation for aqueous laponite suspensions with different salt concentration. Filled symbols represent clay concentration 3.5 % with different molarity, from top to bottom: $10^{-2}$ M, $7\times10^{-3}$ M, $5\times10^{-3}$ M, $3\times10^{-3}$ M, $10^{-3}$ M and $10^{-4}$ M. Open circles represent 2 %, $10^{-2}$ M clay system. Inset shows dependence of power law exponents $x$ (for $\eta^* \sim t_w^x$) on molar concentration of Na$^+$ ions.

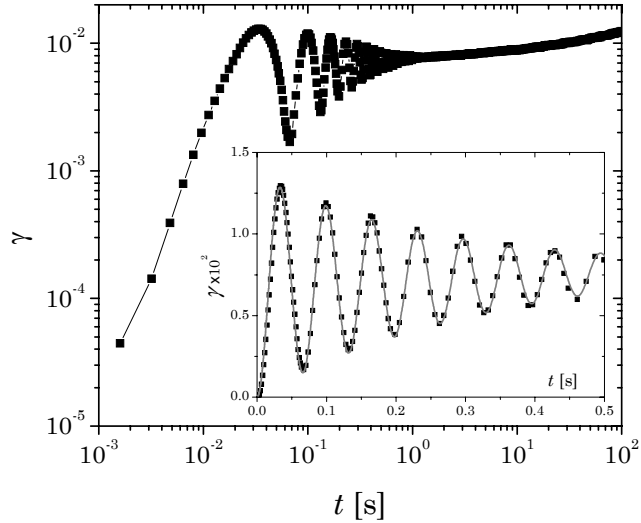

**Figure 4.** Typical creep behavior of aqueous laponite suspension. At early times, damped oscillations in strain are observed. Line is a guide to an eye. Inset shows fit of eq. 3 to the experimental data. It can be seen that only a single mode fits the data very well.

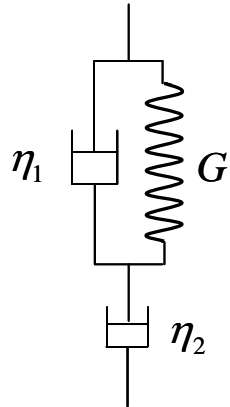

**Figure 5.** Schematic of a Maxwell-Jeffreys model.

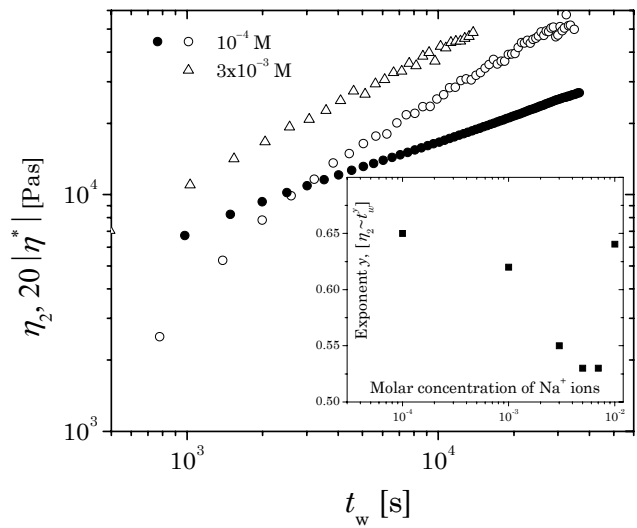

**Figure 6.** Evolution of $\eta_2$ with age after rejuvenation for aqueous laponite suspensions for two salt concentrations. Filled circles shows corresponding evolution of complex viscosity for the system with no salt ($10^{-4}$ M Na$^+$). Inset shows variation of power law exponent for dependence of $\eta_2$ on age for various salt concentrations.

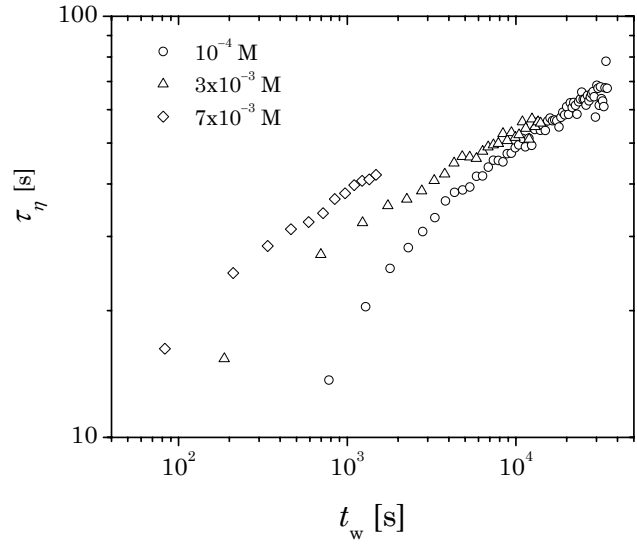

**Figure 7.** Variation of dominating rheological time scale $\tau_\eta$ as a function of age for various salt concentrations.

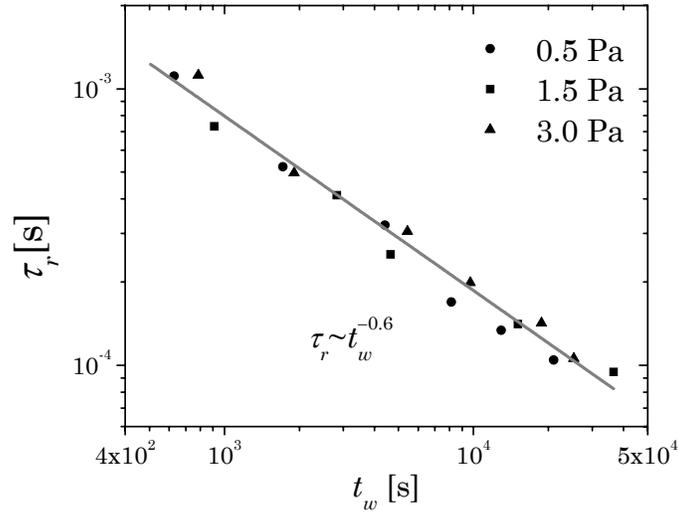

**Figure 8.** Dependence of retardation time $\tau_r$ on age at various stresses for the 3.5 %, $10^{-4}$ M sample.

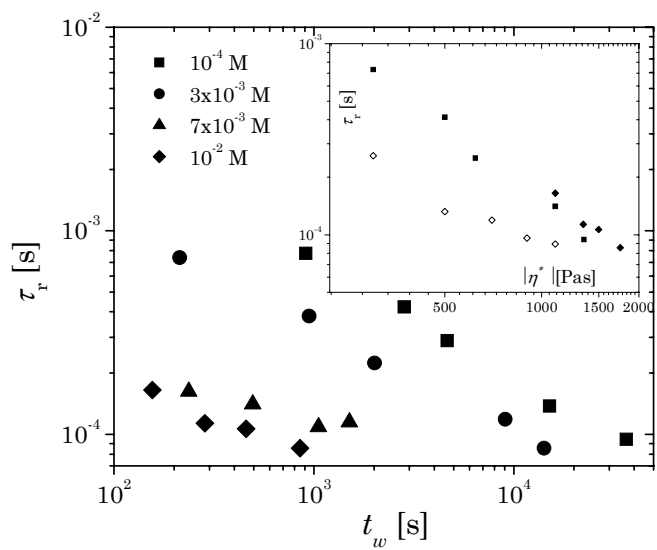

**Figure 9.** Dependence of retardation time $\tau_r$ on age. Inset shows dependence of the same on complex viscosity. Open symbols in the inset represent 2 %, $10^{-2}$ M suspension.

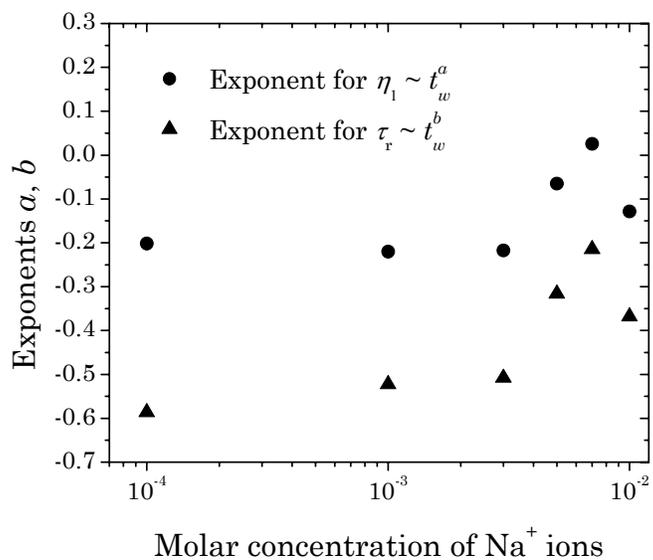

**Figure 10.** Dependence of power law exponents $a$ and $b$ (for $\eta_1 \sim t_w^a$ and $\tau_r \sim t_w^b$ respectively) on molar concentration of $Na^+$ ions.

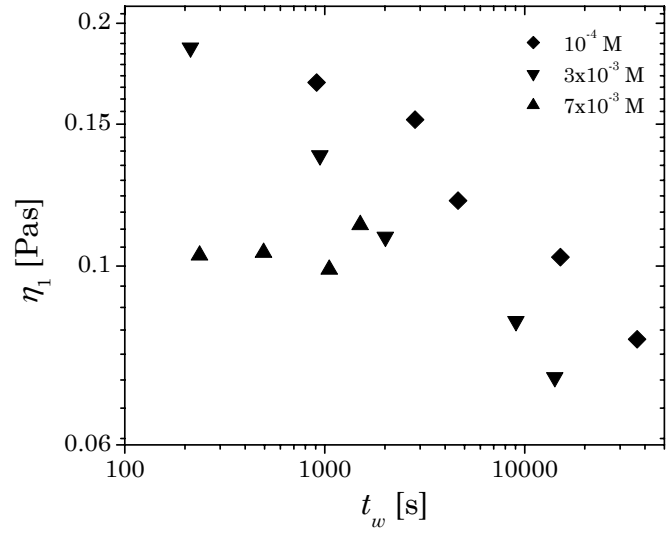

**Figure 11.** Evolution of $\eta_1$ as a function of age for various salt concentrations.

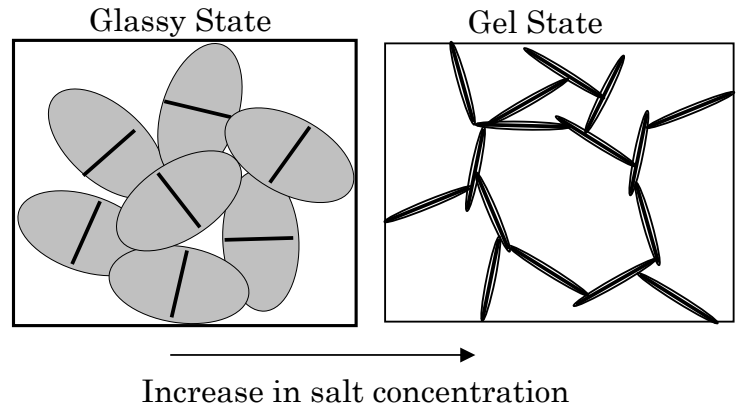

Increase in salt concentration

**Figure 12.** Schematic of arrested states. At low ionic concentration, electrostatic repulsion between laponite particles leads to glassy state. At high ionic concentration, cations screen the negative charge which leads to a gel state.

**References:**

Abou, B., Bonn, D. & Meunier, J. 2001 Aging dynamics in a colloidal glass. *Phys. Rev. E* **64**, 215101-215106.

Abou, B., Bonn, D. & Meunier, J. 2003 Nonlinear rheology of Laponite suspensions under an external drive. *J. Rheol.* **47**, 979-988.

Avery, R. G. & Ramsay, J. D. F. 1986 Colloidal properties of synthetic hectorite clay dispersions. II. Light and small angle neutron scattering. *J.Colloid Interface Sci.* **109**, 448-454.

Baravian, C. & Quemada, D. 1998 Using instrumental inertia in controlled stress rheometry. *Rheologica Acta* **37**, 223-233.

Baravian, C., Vantelon, D. & Thomas, F. 2003 Rheological determination of interaction potential energy for aqueous clay suspensions. *Langmuir* **19**, 8109-8114.

Barnes, H. A. 1997 Thixotropy - A review. *Journal of Non-Newtonian Fluid Mechanics* **70**, 1-33.

Barnes, H. A., Hutton, J. F. & Walters, K. 1989 *An Introduction to Rheology*. Amsterdam: Elsevier.

Bellour, M., Knaebel, A., Harden, J. L., Lequeux, F. & Munch, J.-P. 2003 Aging processes and scale dependence in soft glassy colloidal suspensions. *Phys. Rev. E* **67**, 031405.

Bhatia, S., Barker, J. & Mourchid, A. 2003 Scattering of disklike particle suspensions: Evidence for repulsive interactions and large length scale structure from static light scattering and ultra-small-angle neutron scattering. *Langmuir* **19**, 532-535.

Bonn, D., Coussot, P., Huynh, H. T., Bertrand, F. & Debregeas, G. 2002a Rheology of soft glassy materials. *Europhysics Letters* **59**, 786-792.

Bonn, D., Kellay, H., Tanaka, H., Wegdam, G. & Meunier, J. 1999 Laponite: What is the difference between a gel and a glass? *Langmuir* **15**, 7534-7536.

Bonn, D., Tanaka, H., Wegdam, G., Kellay, H. & Meunier, J. 1998 Aging of a colloidal "Wigner" glass. *Europhys. Lett.* **45**, 52.

Bonn, D., Tanasc, S., Abou, B., Tanaka, H. & Meunier, J. 2002b Laponite: Aging and shear rejuvenation of a colloidal glass. *Phys. Rev. Lett.* **89**, 157011-157014.

Cipelletti, L. & Ramos, L. 2005 Slow dynamics in glassy soft matter. *J. Phys. Con. Mat.* **17**, R253–R285.

Cloitre, M., Borrega, R. & Leibler, L. 2000 Rheological aging and rejuvenation in microgel pastes. *Phys. Rev. Lett.* **85**, 4819-4822.

Cloitre, M., Borrega, R., Monti, F. & Leibler, L. 2003 Glassy dynamics and flow properties of soft colloidal pastes. *Phys. Rev. Lett.* **90**, 068303.

Cocard, S., Tassin, J. F. & Nicolai, T. 2000 Dynamical mechanical properties of gelling colloidal disks. *J. Rheol.* **44**, 585-594.

Cohen-Addad, S. & Hohler, R. 2001 Bubble dynamics relaxation in aqueous foam probed by multispeckle diffusing-wave spectroscopy. *Phys. Rev. Lett.* **86**, 4700-4703.

Coussot, P. 2006 Rheological aspects of the solid-liquid transition in jammed systems. *Lecture Notes in Physics* **688**, 69-90.

Coussot, P. 2007 Rheophysics of pastes: A review of microscopic modelling approaches. *Soft Matter* **3**, 528-540.

Coussot, P., Tabuteau, H., Chateau, X., Tocquer, L. & Ovarlez, G. 2006 Aging and solid or liquid behavior in pastes. *J. Rheol.* **50**, 975-994.

Di Leonardo, R., Ianni, F. & Ruocco, G. 2005 Aging under shear: Structural relaxation of a non-Newtonian fluid. *Phys. Rev. E* **71**, 011505.

Fielding, S. M., Sollich, P. & Cates, M. E. 2000 Aging and rheology in soft materials. *J. Rheol.* **44**, 323-369.

Findley, W. N., Lai, J. S. & Onaran, K. 1976 *Creep and Relaxation of Nonlinear Viscoelastic Materials*. Amsterdam: North-Holland.

Gabriel, J.-C. P., Sanchez, C. & Davidson, P. 1996 Observation of nematic liquid-crystal textures in aqueous gels of smectite clays. *J. Phys. Chem.* **100**, 11139-11143.

Gang, H., Krall, A. H., Cummins, H. Z. & Weitz, D. A. 1999 Emulsion glasses: A dynamic light-scattering study. *Phys. Rev. E* **59**, 715-721.

Harrison, G. 1976 *The Dynamics Properties of Supercooled Liquids*. London: Academic.

Ianni, F., Di Leonardo, R., Gentilini, S. & Ruocco, G. 2007 Aging after shear rejuvenation in a soft glassy colloidal suspension: Evidence for two different regimes. *Phys. Rev. E* **75**, 011408.

Jabbari-Farouji, S., Wegdam, G. H. & Bonn, D. 2007 Gels and glasses in a single system: Evidence for an intricate free-energy landscape of glassy materials. *Phys. Rev. Lett.* **99**, 065701.

Joshi, Y. M. 2007 Model for cage formation in colloidal suspension of laponite. *Journal of Chemical Physics* **127**, 081102.

Joshi, Y. M. & Reddy, G. R. K. 2007 Aging in a colloidal glass in creep flow: A master curve for creep compliance. *arXiv:0710.5264*.

Kaloun, S., Skouri, R., Skouri, M., Munch, J. P. & Schosseler, F. 2005 *Phys. Rev. E* **72**, 011403.

Knaebel, A., Bellour, M., Munch, J.-P., Viasnoff, V., Lequeux, F. & Harden, J. L. 2000 Aging behavior of Laponite clay particle suspensions. *Europhysics Letters* **52**, 73-79.

Kroon, M., Vos, W. L. & Wegdam, G. H. 1998 Structure and formation of a gel of colloidal disks. *Phys. Rev. E* **57**, 1962-1970.

Kroon, M., Wegdam, G. H. & Sprik, R. 1996 Dynamic light scattering studies on the sol-gel transition of a suspension of anisotropic colloidal particles. *Phys. Rev. E* **54**, 6541-6550.

Levitz, P., Lecolier, E., Mourchid, A., Delville, A. & Lyonnard, S. 2000 Liquid-solid transition of Laponite suspensions at very low ionic strength: Long-range electrostatic stabilisation of anisotropic colloids. *Europhysics Letters* **49**, 672-677.

Liu, A. J. & Nagel, S. R. 1998 Jamming is not just cool any more. *Nature* **396**, 21-22.

Luckham, P. F. & Rossi, S. 1999 Colloidal and rheological properties of bentonite suspensions. *Advances in Colloid and Interface Science* **82**, 43-92.

Mallamace, F., Gambadauro, P., Micali, N., Tartaglia, P., Liao, C. & Chen, S.-H. 2000 Kinetic Glass Transition in a Micellar System with Short-Range Attractive Interaction. *Phys. Rev. Lett.* **84**, 5431-5434.

Marin, G. 1988 Oscillatory Rheometry. In *Rheological Measurements* (ed. A. A. Collyer & D. W. Clegg), pp. 297-343. London: Elsevier.

Michot, L. J., Bihannic, I., Maddi, S., Funari, S. S., Baravian, C., Levitz, P. & Davidson, P. 2006 Liquid-crystalline aqueous clay suspensions. *PNAS* **103**, 16101-16104.

Mongondry, P., Tassin, J. F. & Nicolai, T. 2005 Revised state diagram of Laponite dispersions. *J.Colloid Interface Sci.* **283**, 397-405.

Mossa, S., De Michele, C. & Sciortino, F. 2007 Aging in a Laponite colloidal suspension: A Brownian dynamics simulation study. *Journal of Chemical Physics* **126**, 014905.

Mourchid, A., Delville, A., Lambard, J., Lecolier, E. & Levitz, P. 1995a Phase diagram of colloidal dispersions of anisotropic charged particles: Equilibrium properties, structure, and rheology of laponite suspensions. *Langmuir* **11**, 1942-1950.

Mourchid, A., Delville, A. & Levitz, P. 1995b Sol-gel transition of colloidal suspensions of anisotropic particles of laponite. *Faraday Discuss.* **101**, 275-285.

Nicolai, T. & Cocard, S. 2000 Light scattering study of the dispersion of laponite. *Langmuir* **16**, 8189-8193.

Nicolai, T. & Cocard, S. 2001 Structure of gels and aggregates of disk-like colloids. *European Physical Journal E* **5**, 221-227.

Ovarlez, G. & Coussot, P. 2007 Physical age of soft-jammed systems. *Phys. Rev. E* **76**, 011406.

Pignon, F., Magnin, A. & Piau, J.-M. 1997a Butterfly light scattering pattern and rheology of a sheared thixotropic clay gel. *Phys. Rev. Lett.* **79**, 4689-4692.

Pignon, F., Magnin, A., Piau, J.-M., Cabane, B., Lindner, P. & Diat, O. 1997b Yield stress thixotropic clay suspension: Investigations of structure by light, neutron, and x-ray scattering. *Phys. Rev. E* **56**, 3281-3289.

Pignon, F., Piau, J.-M. & Magnin, A. 1996 Structure and pertinent length scale of a discotic clay gel. *Phys. Rev. Lett.* **76**, 4857-4860.

Ravi Kumar, N. V. N., Muralidhar, K. & Joshi, Y. M. 2007 On Refractive Index of Ageing Suspensions of Laponite. *submitted to Appl.Clay Sci.*

Ruzicka, B., Zulian, L. & Ruocco, G. 2004 Ergodic to non-ergodic transition in low concentration, Laponite. *J. Phys. Con. Mat.* **16**, S4993.

Ruzicka, B., Zulian, L. & Ruocco, G. 2006 More on the phase diagram of laponite. *Langmuir* **22**, 1106-1111.

Ruzicka, B., Zulian, L. & Ruocco, G. 2007 Ageing dynamics in Laponite dispersions at various salt concentrations. *Philosophical Magazine* **87**, 449-458.

Schosseler, F., Kaloun, S., Skouri, M. & Munch, J. P. 2006 *Phys. Rev. E* **73**, 021401.

Sciortino, F. & Tartaglia, P. 2005 Glassy colloidal systems. *Advances in Physics* **54**, 471-524.

Struik, L. C. E. 1978 *Physical Aging in Amorphous Polymers and Other Materials*. Houston: Elsevier.
Tanaka, H., Jabbari-Farouji, S., Meunier, J. & Bonn, D. 2005 Kinetics of ergodic-to-nonergodic transitions in charged colloidal suspensions: Aging and gelation. *Phys. Rev. E* **71**, 021402.
Tanaka, H., Meunier, J. & Bonn, D. 2004 Nonergodic states of charged colloidal suspensions: Repulsive and attractive glasses and gels. *Phys. Rev. E* **69**, 031404.
Treece, M. A. & Oberhauser, J. P. 2007 Soft glassy dynamics in polypropylene-clay nanocomposites. *Macromolecules* **40**, 571-582.
Utz, M., Debenedetti, P. G. & Stillinger, F. H. 2000 Atomistic Simulation of Aging and Rejuvenation in Glasses. *Phys. Rev. Lett.* **84**, 1471-1474.
Van Olphen, H. 1977 *An Introduction to Clay Colloid Chemistry*. New York: Wiley.
Viasnoff, V. & Lequeux, F. 2002 Rejuvenation and overaging in a colloidal glass under shear. *Phys. Rev. Lett.* **89**, 065701.
Weeks, E. R., Crocker, J. C. & Weitz, D. A. 2007 Short- and long-range correlated motion observed in colloidal glasses and liquids. *J. Phys. Con. Mat.* **19**, 205131.